\documentclass[aps,prl,amsmath]{revtex4}
\usepackage{graphicx}
\usepackage{color}
\usepackage{bm}
\usepackage{epsfig}
\usepackage{latexsym}
\usepackage{pifont}
\usepackage{float}
\usepackage[utf8]{inputenc}
\graphicspath{{figure/}}
\usepackage{siunitx}
\usepackage{textgreek}

\begin{document}
\newcommand{\be}{\begin{equation}}
\newcommand{\ee}{\end{equation}}
\newcommand{\bea}{\begin{eqnarray}}
\newcommand{\eea}{\end{eqnarray}}
\newcommand{\nn}{\nonumber}

\title{Coiling of cellular protrusions around extracellular fibers}

\author{Raj Kumar Sadhu$^1$, Christian Hernandez-Padilla$^2$, Yael Eshed Eisenbach$^3$, Lixia Zhang$^4$, Harshad D Vishwasrao$^4$, Bahareh Behkam$^2$,  Hari Shroff$^{4,5}$, Ale\v{s} Igli\v{c}$^6$, Elior Peles$^3$, Amrinder S. Nain$^2$ and Nir S Gov$^1$ }
\affiliation{$^1$Department of Chemical and Biological Physics, Weizmann Institute of Science, Rehovot 7610001, Israel; $^2$Department of Mechanical Engineering, Virginia Tech, Blacksburg, VA 24061, USA; $^3$Department of Molecular Cell Biology, Weizmann Institute of Science, Rehovot 7610001, Israel; $^4$Advanced Imaging and Microscopy Resource, National Institutes of Health, Bethesda, Maryland, USA; $^5$Laboratory of High Resolution Optical Imaging, National Institute of Biomedical Imaging and Bioengineering, National Institutes of Health, Bethesda, Maryland, USA; $^6$Laboratory of Physics, Faculty of Electrical Engineering, University of Ljubljana, Ljubljana, Slovenia}

\begin{abstract}
Protrusions at the leading-edge of a cell play an important role in sensing the extracellular cues, during cellular spreading and motility. Recent studies provided indications that these protrusions wrap (coil) around the extra-cellular fibers. The details of this coiling process, and the mechanisms that drive it, are not well understood. We present a combined theoretical and experimental study of the coiling of cellular protrusions on fibers of different geometry. Our theoretical model describes membrane protrusions that are produced by curved membrane proteins that recruit the protrusive forces of actin polymerization, and identifies the role of bending and adhesion energies in orienting the leading-edges of the protrusions along the azimuthal (coiling) direction. Our model predicts that the cell's leading-edge coils on round fibers, but the coiling ceases for a fiber of elliptical (flat) cross-section. These predictions are verified by 3D visualization and quantitation of coiling on suspended fibers using Dual-View light-sheet microscopy (diSPIM). Overall, we provide a theoretical framework supported by high spatiotemporal resolution experiments capable of resolving coiling of cellular protrusions around extracellular fibers of varying diameters. 
\end{abstract}

\maketitle

\section{Introduction}\label{sen:intro}
Cellular protrusions play important roles in exploring and sensing the extracellular environment, during cell spreading and adhesion, cell migration, and cell-cell interaction \cite{review1,review2,review3,Carlson2009,Veranic2008}. Lamellipodia and filopodia are protrusive structures formed at the leading-edge of a migratory cell \cite{RIDLEY2011,Friedl2009,EILKEN2010,Charras2008,Buccione2009,diz-munoj2010}. These protrusions enable cells to adhere and spread on fiber-like surfaces \cite{Sebastien2020,Richard2019,koons2017}, such as the fibers of the extra-cellular matrix (ECM) \cite{CLARK1982264,ushiki2002}, as well as cylindrical protrusions of other cells, such as glial cells spreading over neighboring axonal extensions \cite{Christine2019}. \textit{In vitro} studies of the cellular spreading and migration on fibers \cite{Sebastien2020,Richard2019} have shown how different cell types organize on these fibers \cite{Nathan2017,Svitkina1995,Lee2012,Hwang2009,Werner2018,Werner2019}, with the cellular shape and motility found to depend on the curvature (diameter) of the fibers \cite{nain2014,KENNEDY201741,Nathan2017,Lee2012,Hwang2009,Werner2018,Werner2019}.

\begin{figure}[h!]
\centering
\includegraphics[scale=0.18]{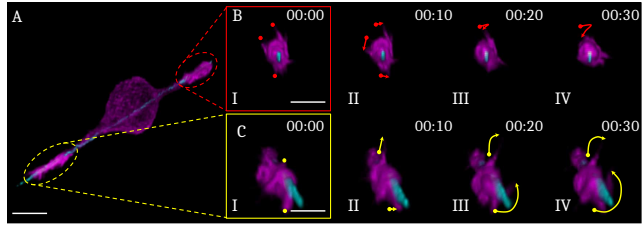}
\caption{Coiling events occurring at the leading-edges of cellular protrusions while suspended on a single fiber of $200nm$ diameter. (A) 3D render of a C2C12 cell captured with Dual-View Light Sheet microscopy. C2C12 cells were tagged with GFP actin (shown in Magenta), and the polystyrene fiber was coated with Rhodamine Fibronectin (shown in Cyan). Scale bar is 10 $\mu$m (B) I-IV time-lapse sequence images of cellular protrusion highlighted by the red oval in A. Images track with arrows (frames II-IV) the path of the membrane ruffles marked with dots in the frame I, during 10-second intervals. (C) I-IV time-lapse sequence images that follow the same description as in B, but for the leading-edge highlighted by the yellow oval in A. The scale bar shown in frame I is 5 $\mu m$.}
\label{fig:coiling_example}
\end{figure}

Experiments studying the membrane dynamics at the leading-edge of cellular protrusions, have found indications for coiling (wrapping) dynamics around the extracellular fibers. In \cite{mukherjee2019cancer,nain2022}, protrusions of metastatic cancer cells (breast and ovarian) were observed to coil and rotate around the fiber's axis in a curvature-dependent manner, while in \cite{Nils2015} similar coiling dynamics of leading-edge `fin'-like protrusions were observed for several cell types (fibroblasts, epithelial, endothelial). These protrusions are important during the cell's adhesion and spreading, and play an important role in maintaining the cell polarity and migration. 

The mechanism that gives rise to the rotational spreading of the membrane on the fibers is not understood at present. Here we use a theoretical model \cite{Sadhu2021} where the leading-edge protrusion forms due to forces produced by curved actin nucleators, combined with adhesion to the substrate. Curved membrane proteins that recruit the actin cytoskeleton have been recently located at the leading edge of lamellipodia protrusions \cite{Orion2021,Galic2019}. We show here that this model spontaneously gives rise to the coiling motion of the membrane when the vesicle spreads over a cylindrical surface that represents the external fiber, and therefore may offer a mechanism for the observed coiling dynamics in cells. Motivated by this theoretical result, we use Dual-View light-sheet microscopy (diSPIM) \cite{Wu2013,Kumar2014} to obtain high-resolution imaging of these coiling protrusions (Fig.\ref{fig:coiling_example}, Movie-S1-S3), which we compare to the theoretical predictions.
%%%%%%%%%%%%%%%%%%%%%%%%%%%%%%%%%%%%%%%%%%%%%%%%%%%%%%%%%%%%%%%%%%%%%%%%%%%%%%%%%%%%%%%%%%%%%%%%%%%%%

\section{Results}\label{sec:result}
\begin{figure*}[h!]
\centering
\includegraphics[scale=0.7]{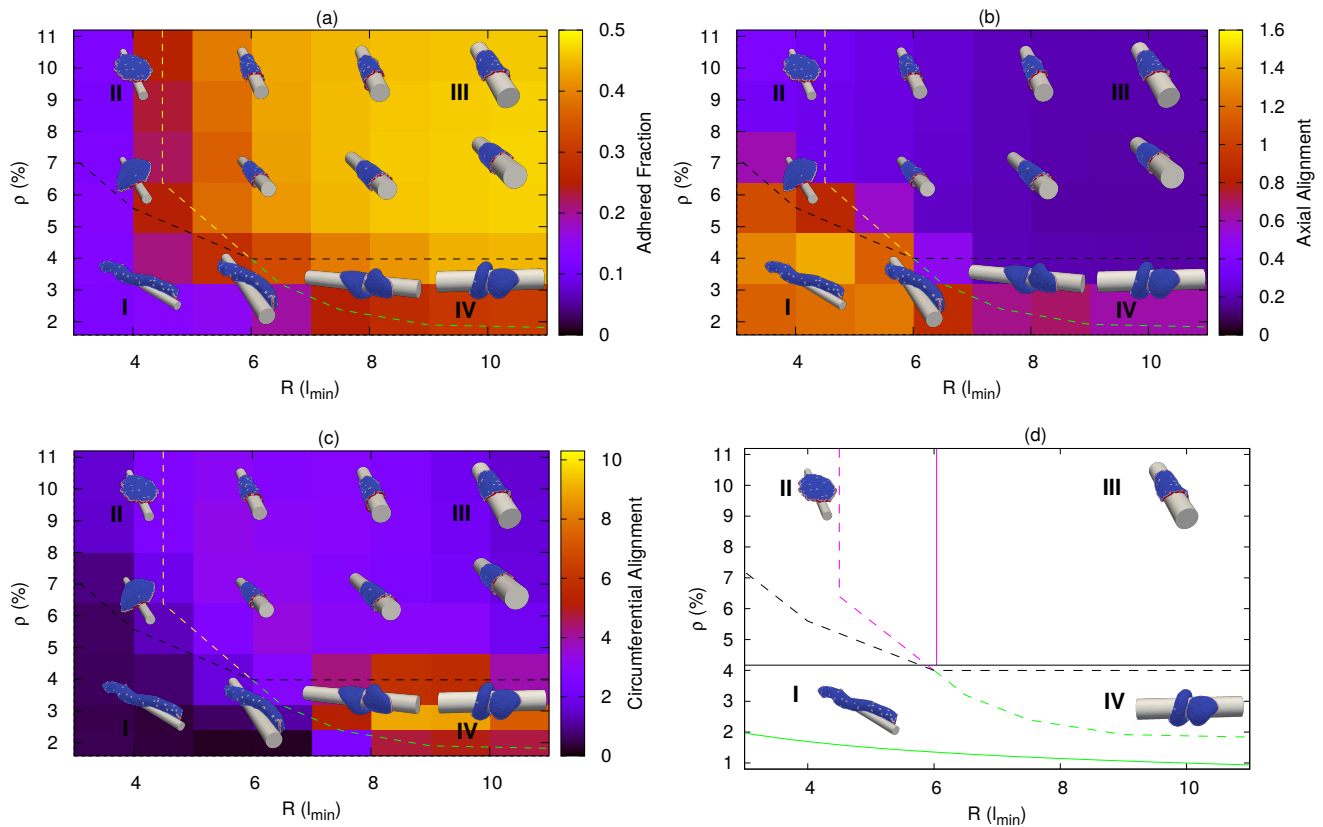}
\caption{Phase-diagram of the steady-state vesicle shapes in the $R-\rho$ plane, where $R$ is the fiber radius and $\rho$ is the density of curved proteins. The snapshots are shown for $R=4, 6, 8, 10$ (in units of $l_{min}$), and $\rho=3.2 \%, 6.4 \%$ and $9.6 \%$. The red regions in the snapshots denote the curved proteins while the blue regions denote the bare membrane. Here, we use adhesion strength $(E_{ad})=1.0 k_BT$, $F=2.0 k_BT/l_{min}$. (a) The background color is showing the adhered area fraction of the vesicle, which is maximal for large $R$ and $\rho$ (phase III). (b) The background color is showing the variance of the vertices locations along the cylinder axis (scaled by $10^3~l_{min}^2$), which is maximal for small $R$ and $\rho$, where the vesicle is aligned along the axis (phase I). (c) The background color is showing the angular variance of the vertices along the circumferential direction, which is maximal when the vesicle coils over the cylinder for large $R$ and small $\rho$ (phase IV). (d) Comparison between the transition lines between the different vesicle shapes from the simulations (dashed lines) and the analytical predictions (solid lines).}
\label{fig:active}
\end{figure*}

In our theoretical model, we consider a three-dimensional cell-like vesicle,  which is described by a closed surface, having $N$ vertices, each of them connected to its neighbors with bonds, forming a dynamically triangulated, self-avoiding network, with the topology of a sphere \cite{miha2019,Sadhu2021,Sadhu_phagocytosis}. The membrane contains proteins with convex spontaneous curvature, that diffuse on the surface of the vesicle, having attractive nearest neighbour  interactions with each other. Each curved protein exerts a force ($F$) in the direction of the local outward normal of the surface, representing the protrusive force due to actin polymerization (see the Methods section for details).

%%%%%%%%%%%%%%%%%%%%%%%%%%%%%%%%%%%%%%%%%%%%%%%%%%%%%%%%%%%%%%%%%%%%%%%%%%%%%%%%%%%%%%%%%%%%%%%%%%%%
\subsection{Shapes of vesicles spreading on the fiber}
We start by analyzing the dynamics of how a vesicle spreads on an adhesive fiber of a circular cross-section. Spreading of the vesicle over an adhesive cylindrical surface is determined by the balance between the bending and adhesion energies \cite{Sadhu2021}. The curved membrane proteins, even when passive (do not recruit the protrusive forces due to actin polymerization, $F=0$), can enhance the spreading by reducing the bending energy cost. This is shown in SI sec. S1, Figs. S1, S2 (Movie-S4-S8), with a monotonously increasing adhered area as the adhesion energy, radius of the cylindrical fiber ($R$) and the average density of curved proteins ($\rho$) increase.  These systems do not exhibit any tendency for rotations or coiling dynamics.

In Fig.\ref{fig:active} we describe the steady-state shapes for vesicles with active curved proteins ($F \neq 0$), as a function of $R$ and $\rho$. In Fig. \ref{fig:active}(a), we plot the adhered fraction of the vesicle (background color), along with the steady-state snapshots. We notice that there are several distinct phases of adhered vesicles on the fiber, which are marked by the colored ``transition" lines (Fig.\ref{fig:active}d). The details of the analytic calculation of the transition lines are given in the SI Sec. S2.

I) For small $R$ and $\rho$, the vesicle shape is elongated (aligned axially, along the long axis of the fiber), with a very small adhered area. Since $R$ is small, and there are not enough curved proteins to form a ring-like aggregate around the whole rim of the vesicle (as in phase II, above), the vesicle can only adhere axially and in this way minimize its bending energy. We identify this phase as it maximizes the axial elongation of the adhered vesicle (Fig.\ref{fig:active}b, Movie-S9) as measured by the variance in the distribution of vertices along the axial direction (Z-axis) (background color in Fig. \ref{fig:active}b, and SI sec. S3, Fig. S4).

II) For small $R$, if we increase $\rho$, the vesicle will eventually form a flat pancake-like shape with all the proteins clustered as a ring around the rim. The protrusive forces in this case act side-way around the ring and make this pancake-like shape stable. These type of shapes were also observed for a free vesicle (without any adhesive substrate) \cite{miha2019}, when the protein density $\rho$ reaches a critical value. For small $R$ the bending energy cost of pancake-like vesicle for wrapping around the fiber is too high, and it remains ``hanging" on the fiber \cite{koons2017} (Movie-S10). 

III) At large $\rho$ (such that the vesicle is able to form a pancake-like shape), if we increase $R$, the vesicle will fully adhere, as the adhesion energy gain now dominates over the bending energy cost of wrapping around the fiber. This phase is identified by having the maximal adhered area fraction of $\sim 0.50$ (background color in Fig.\ref{fig:active}a, Movie-S11). 

IV) At large $R$, if $\rho$ is small enough so that the proteins can not form a circular aggregate around the flat (pancake-like) vesicle, they form a two-arc phase \cite{miha2019,Sadhu2021}, where two aggregates of proteins form at opposing ends of the cell, stretching the elongated membrane between them. This shape spontaneously orients to point along the circumferential direction, and pull the vesicle into a coiled helical structure. This phase is therefore identified by the large overall angular spread of the vesicle along its length (Fig.\ref{fig:active}c, Movie-S12), quantified by the variance in the angular distribution of the vertices along the circumferential direction (background color in Fig. \ref{fig:active}(c) and SI sec. S3, Fig. S5).

Note that in the parameter regime of phase IV, the vesicles can also form a phenotype where all the proteins are aggregated in a single cluster. These vesicles become motile, as observed on a flat adhesive substrate \cite{Sadhu2021}.  

%%%%%%%%%%%%%%%%%%%%%%%%%%%%%%%%%%%%%%%%%%%%%%%%%%%%%%%%%%%%%%%%%%%%%%%%%%%%%%%%%%%%%%%%%%%%%%%%%%%%

\subsection{The mechanism driving the coiling phase}

We now analyze the coiling phase (IV, Fig. \ref{fig:active}), to expose the mechanism that drives the circumferential orientation of the leading-edges of the membrane protrusions. 

\begin{figure*}[h!]
\centering
\includegraphics[scale=1.13]{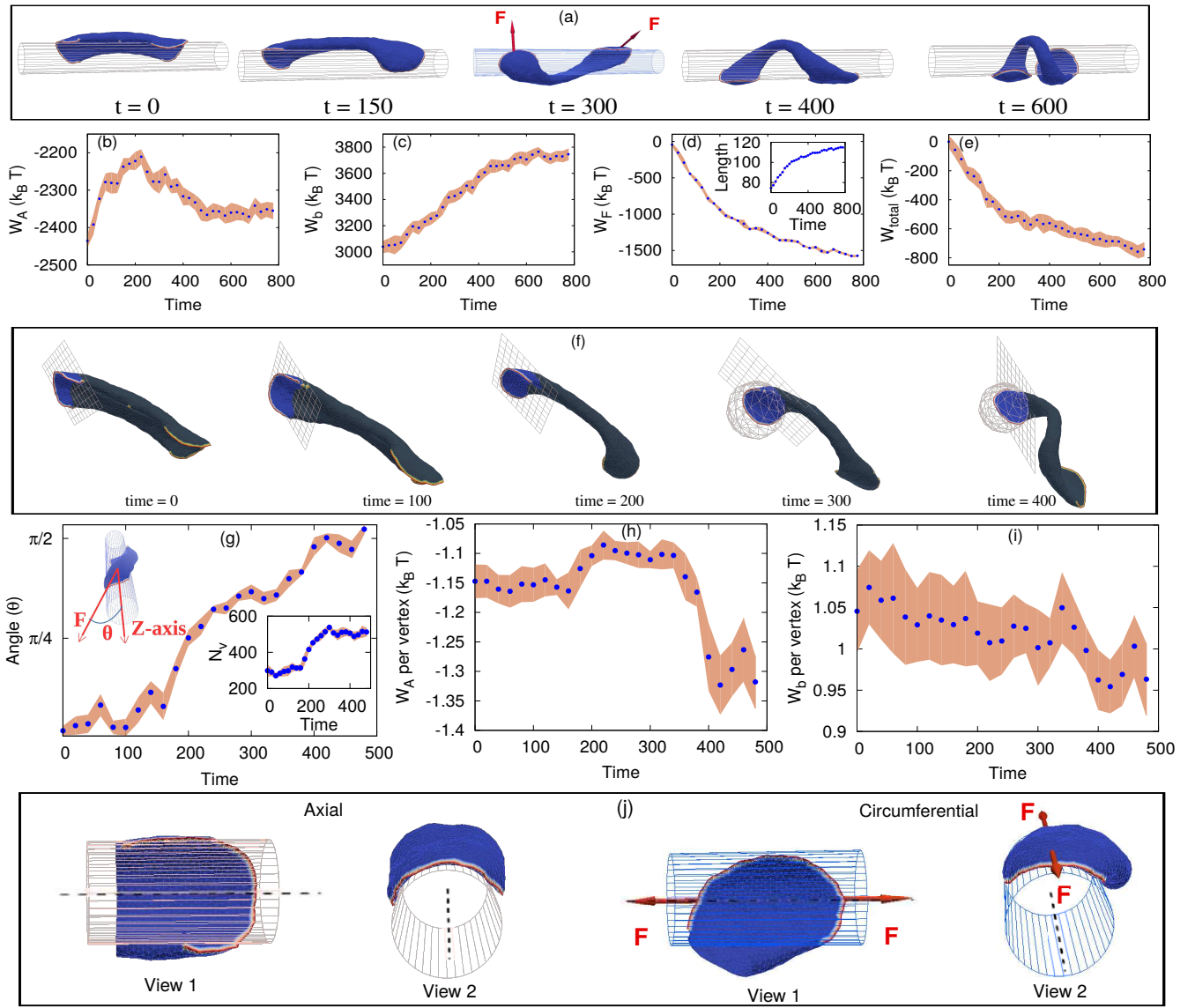}
\caption{Transition of a two arc shape from axial to circumferential (coiling) orientation. (a) Configurations of the vesicle at different times (in Monte-Carlo units). The red arrows on the 3rd inset are showing the direction of active forces acting on each of the two arcs of the elongated vesicle. (b) Variation of the total adhesion energy ($W_A$) with time, showing a non-monotonic variation. (c) The total bending energy ($W_b$) of the vesicle as a function of time, which is increasing monotonously during the coiling process. (d) Total work done by the active force ($W_F$), increasing the vesicle length over time (shown in the inset). The length is measured by assuming that the vesicle is organized like a helix, with the end positions being the centers of mass of the proteins in each of the two leading-edge arcs. (e) The total effective energy ($W_{total}$), sum of b-d, which is a decreasing function of time. (f) Configurations showing the edge region, as defined by the grid-plane, and the grid-sphere at later times when the shape is highly coiled. (g) The angle ($\theta$) between the direction of the total force acting at the cell edge (as denoted in (f)), with the cylindrical axis as a function of time. The inset is showing the total number of vertices within this region. (h) Adhesion energy and (i) bending energy per vertex for the leading-edge region (f). (j) The configuration of an arc at the leading-edge when it is oriented axially (as in $t=0$ in (a)), versus circumferentially (as in $t=600$ in (a)). The arrows denote the active axial stretching when the leading-edge is oriented circumferentially, increasing the adhesion energy (h).  Here, we use $E_{ad}=2.0 k_BT$, $\rho=2.4 \%$ and $F=2.0 k_BT/l_{min}$. The unit of time in the plots is normalized to be $2 \times 10^4$ MC steps.}
\label{coiling_time}
\end{figure*}

We consider a two-arc vesicle, generated on a flat substrate, and place it on the cylindrical fiber along the axial orientation as shown in Fig. \ref{coiling_time}(a). We find that the vesicle is unstable in the axially aligned state, and the two leading-edges spontaneously rotate to circumferential alignment, thus causing the vesicle to coil (Movie-S13). We follow the evolution of the energy components during this process (Fig. \ref{coiling_time}(b-e)). Globally, the adhesion energy ($W_A$) decreases (smaller in absolute value) at the early stage (Fig. \ref{coiling_time}(b), time $ \lesssim 200$), as the tubular middle part of the vesicle partially detaches from the curved fiber surface and is stretched by the active force (Fig. \ref{coiling_time}(d), inset). The overall bending energy ($W_b$) increases throughout the process, as the tubular part is stretched to become thinner and is bent (coiled) circumferentially around the fiber (Fig.\ref{coiling_time}c). 

These energy penalties are counter-balanced by the work done by the active force ($W_F$) in stretching the vesicle (Fig.\ref{coiling_time}d). This energy is calculated as the integrated change in length of the vesicle multiplied by the net active force component that is aligned along the stretch direction (red arrows in Fig.\ref{coiling_time}a at $t=300$). As was shown for the two-arc configuration on a flat substrate \cite{Sadhu2021}, the active work contributes a negative term to the total effective energy of this configuration (Fig.\ref{coiling_time}e), which is the sum $W_{total}=W_A+W_b+W_F$, and acts to stabilize it. At a later stage (time $\gtrsim 200$, Fig.\ref{coiling_time}b), the adhesion energy is partially recovered as the flat leading-edges are stretched along the axial directions by the active forces, increasing their adhered area (Fig. \ref{coiling_time}j, \ref{coiling_time}h). Overall, the total effective energy decreases (becomes more negative) over time (Fig. \ref{coiling_time}e), so the process continues until full coiling.

Note that there is a low probability for the two arcs of the initially axial vesicle (Fig. \ref{coiling_time}a, $t=0$) to reorient in the same direction along the circumference. In such a case, the two arcs will merge to form a motile crescent vesicle \cite{Sadhu_migration}.

While the coiling process is therefore mainly driven by the active forces that elongate the vesicle, performing active work, the above analysis does not explain the origin of the initial reorientation of the leading-edges from the axial to the circumferential alignment. In order to understand this stage, we need to ``zoom-in" on the dynamics of the leading-edges (during the process shown in Fig.\ref{coiling_time}a). We consider a section of the vesicle which contains the leading-edge protein aggregate and the flat membrane protrusion that it forms (Fig.\ref{coiling_time}f). We define this leading-edge region as follows: we draw a plane perpendicular to the direction of the net force of each protein arc, and place it in a position such that all the proteins forming that arc are on one side of this plane. This criteria is not sufficient when the shape of the arc is highly coiled, so we use another constraint (Fig.\ref{coiling_time}f at time$\gtrsim300$): we consider the leading-edge region for all nodes that are within a distance $r_{max}$ from the center-of-mass (COM) of the proteins forming the arc, where $r_{max}$ is the maximum distance of a protein in the arc from the COM of the proteins. These criteria define a leading-edge membrane region that slightly fluctuates in size over time (Fig. \ref{coiling_time}(g) inset, where we see that the biggest change in the number of vertices occurs at time$\sim 300$).

At the beginning of the process the total force due to the proteins in each arc is directed along the axial direction, and the angle of this force with the cylinder's axis increases until it is perpendicular ($\sim \pi/2$) at later times (Fig.\ref{coiling_time}g). The adhesion and bending energies per vertex of the membrane within the leading-edge region are shown in Fig. \ref{coiling_time}(h,i). We plot the energies per vertex to remove the effect of the variation in the number of vertices over time (Fig.\ref{coiling_time}g, inset). The adhesion energy fluctuates in the beginning but there is an overall increase, while the bending energy slightly improves throughout the process. The total value of adhesion and bending energies without scaling by the number of vertices are shown in SI sec. S4, Fig. S6.

These changes arise from the flat membrane at the leading-edge region being more stretched along the axis (the zero curvature direction) by the active proteins when oriented circumferentially, thereby increasing the adhesion energy and reducing the bending energy (Fig. \ref{coiling_time}j). In the initial axial orientation the active forces of the proteins along the leading-edge are not as effective in stretching the membrane sideways, as this involves strong bending of the membrane around the fiber (Fig. \ref{coiling_time}j), and therefore encounters a large bending energy penalty. From these observations we conclude that the reorientation process of the leading-edge regions is mainly driven by locally increasing the adhesion and decreasing the bending energies, while the global coiling process of the whole vesicle is driven by the work done by the active forces. 

%%%%%%%%%%%%%%%%%%%%%%%%%%%%%%%%%%%%%%%%%%%%%%%%%%%%%%%%%%%%%%%%%%%%%%%%%%%%%%%%%%%%%%%%%%%%%%%%%%%%

\subsection{Theoretical predictions compared with experiments}

We now use our theoretical model to make several predictions that we then test in experiments. 

\begin{figure}[h!]
\centering
\includegraphics[scale=0.33]{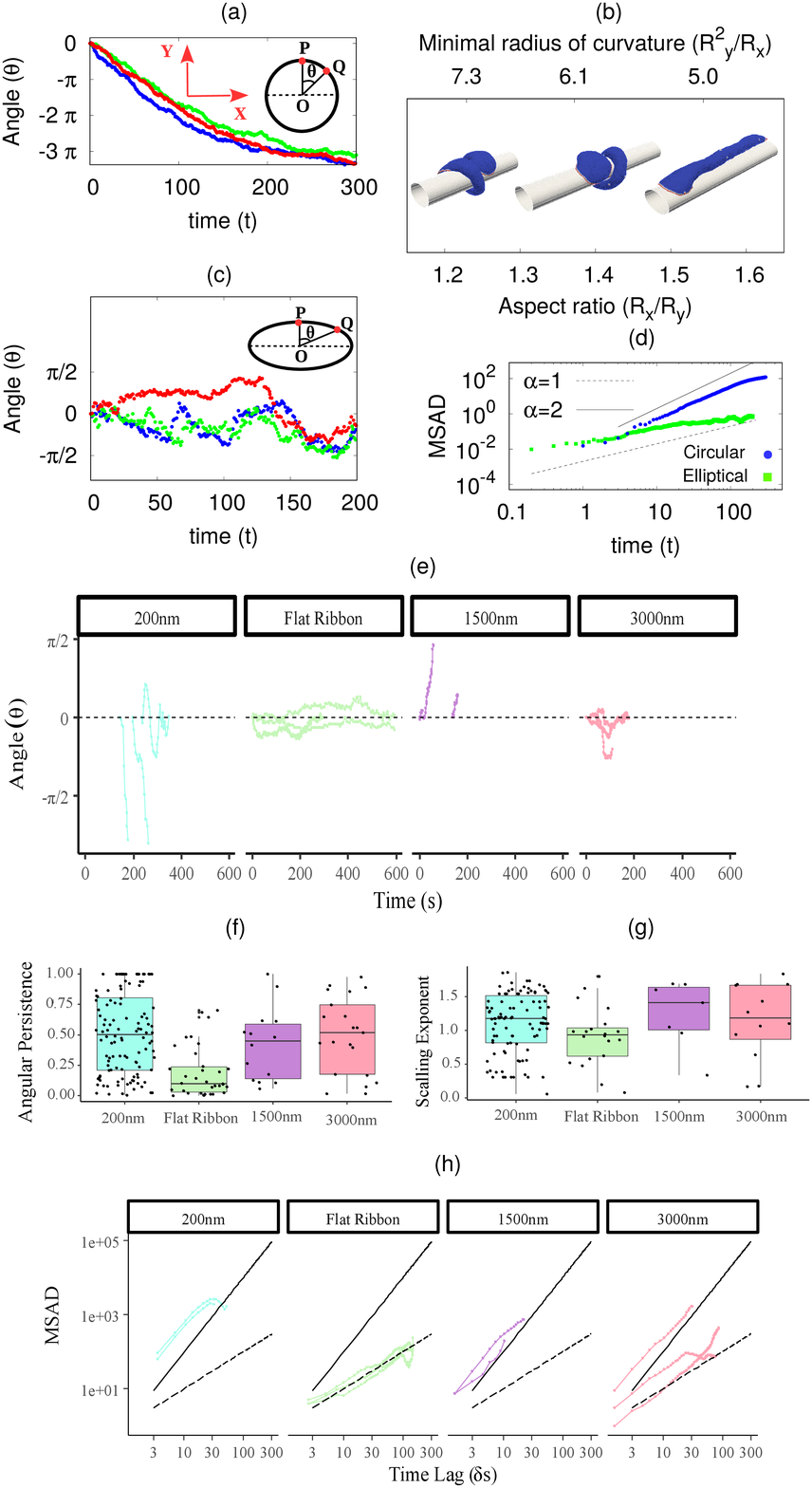}
\caption{Membrane coiling dynamics on fibers of circular and elliptical cross-sections. (a) Angular displacement ($\theta$) of the leading-edge of vesicle with time for the circular fiber ($R=10 l_{min}$). Different colors represent different realizations. The inset shows the definition of angular displacement $\theta$. Here, $P$ and $Q$ are the initial and final position of a leading-edge protein on the $X-Y$ plane, and $\theta$ is the angular displacement between them. We use here $E_{ad}=1.0 k_BT$, $\rho=2.4 \%$ and $F=2.0 k_BT/l_{min}$. The unit of time is $10^5$ MC steps. (b) Configurations of vesicle with fibers of elliptical cross section. The coiling ceases as the aspect ratio of the elliptical cross section ($R_x/R_y$) increase. The circumference of the elliptical cross section is kept constant, which is equal to the circumference of the circular cross-section ($=2 \pi R$) with $R=10.0 l_{min}$. (c) Angular displacement of the leading-edge of the vesicle as function of time, for an elliptical fiber of aspect ratio $1.6$.  Different colors represent different realizations. (d) Mean square angular displacement (MSAD) with time for circular and elliptical fibers. The initial growth of MSAD is $\sim t^\alpha$, where $\alpha=1$ represents diffusive behavior while $\alpha=2$ shows ballistic nature. (e) Experimental data for the angular position with time for different types of fibers. (f) Experimental results for angular persistence for various types of circular and flat fibers. (g) The value of power law exponent ($\alpha$) for the experimental data for various fibers. (h) MSAD for the experimental case for various types of fibers. }
\label{fig:compare}
\end{figure}

In Fig.\ref{fig:compare}a we plot the angular displacement ($\theta$) of the proteins on the leading-edges of the membrane protrusions, where $\theta$ is defined as the angle between the initial and subsequent location of a leading-edge protein (measured from the center of the fiber) on the $X-Y$ plane. We next vary the cross-section shape of the fiber by considering an elliptical cross section having the same circumference as the circular fiber with a radius $R=10 l_{min}$. We find that above a critical aspect ratio (ratio of the semi-major to the semi-minor axis of the ellipse $\sim 1.6 $), the vesicle does not coil but rather remains axially aligned (Fig. \ref{fig:compare}b, Movie-S14). The minimal radius of curvature for this cross-section is $5~l_{min}$, which is a little bit smaller than the radius at which coiling stops for a fiber with a circular cross-section ($R \lesssim 7~l_{min}$ in Fig. \ref{fig:active}). The coiling dynamics of the leading-edges of the vesicle on a fiber with such a high aspect ratio is inhibited, and the value of the angular position of the leading-edge fluctuates around zero  (Fig.\ref{fig:compare}c), while on fibers with circular cross-section the angular displacement of the leading-edges increases beyond $|2\pi|$, and saturates when the vesicle fully coiled  (Fig.\ref{fig:compare}a). 

\begin{figure}[h!]
\centering
\includegraphics[scale=0.175]{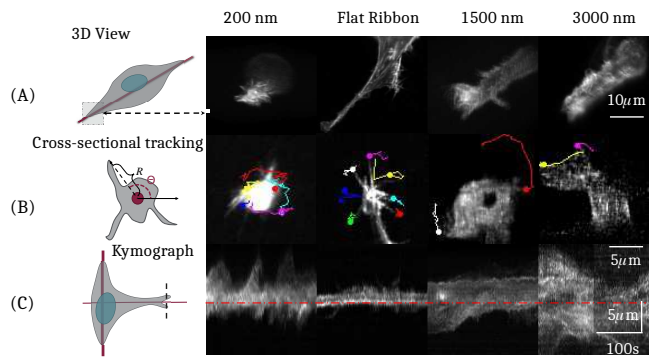}
\caption{Coiling at leading-edges for cells spread on fibers with different diameters. The round polystyrene fibers have diameter of $200$nm, $1500$nm and $3000$nm, while the “Flat Ribbon” fibers are produced by flattening a $200nm$ fiber, increasing its cross-sectional
width to approximately $\pi~ D ~\approx ~600~ nm$ with a thickness $t~ \approx ~50 ~nm$. (A) Maximum intensity projections of the inclined volumetric scan of the cell with Dual-View Light Sheet microscopy. (B) Typical coiling trajectories of membrane ruffles, extracted by cross-sectional manual tracking. The color lines denote the displacements of the tracked ruffle, with the trajectories ending with the dot. (C) Kymographs for the ruffles shown in B. These are illustrating the movement of the membrane boundary along a line placed at the leading-edge of the cell as shown in the dashed line of the illustration on the left.}
\label{fig:experiment}
\end{figure}

We verify this prediction using our experimental results of the trajectories of membrane ruffles for round and flat fibers (experimental images are shown in Fig. \ref{fig:experiment}; details of the experimental methods are in the Material and Method section). Note that in the simulations we model the coiling of the leading-edge that is adhered to the fiber surface, while in the experiments we could visualize the motion of membrane ruffles that extend above the fiber surface. Nevertheless, the motion of these ruffles appears to be correlated with and reflects the qualitative features of the motion of the underlying leading-edge membrane. We find that on a fiber with a round cross-section (diameter $200$ nm) the ruffles exhibit highly directed angular trajectories (Fig.\ref{fig:compare}e), while on the flat-ribbon fibers the trajectories fluctuate around zero (see corresponding trajectories on the cell images in Fig. \ref{fig:experiment}b), in close agreement with the theoretical simulations (Fig.\ref{fig:compare}a,c). The directed trajectories persist for fibers of larger diameter ($1500$nm), but lose their directionality when the fiber diameter is even larger  ($3000$nm). These differences in the experimental trajectories can also be quantified by their angular persistence (Fig. \ref{fig:compare}f). 

The difference between coiling dynamics on round and flat fibers can be further quantified by plotting the  mean square angular displacement (MSAD) for both cases. The MSAD is expected to vary as $\sim t^{\alpha}$, where $\alpha=1$ for a diffusive behavior, while $\alpha=2$ represents ballistic, persistent coiling. In the simulations we find ballistic behavior for round fibers (before the coiling saturates due to finite membrane area), while diffusive for fibers with highly elliptic cross-section (Fig.\ref{fig:compare}d). The experimental data exhibits the same trends (Fig.\ref{fig:compare}h), with ballistic dynamics on the round fibers (of diameter $200$ nm and $1500$ nm), and diffusive motion on the flat-ribbon fiber. On the round fibers with the largest diameter ($3000$nm), we observe mixed behavior. The values of the exponent $\alpha$, extracted from the experimental data, are summarized in Fig.\ref{fig:compare}g. The coiling velocity for each of the experiments is shown in SI sec. S5, Fig. S7, compared with the simulations.

The experimental images of the cells' leading-edges and the trajectories of the ruffles are shown in Fig. \ref{fig:experiment} (also see Movie-S15-S19). Fig. \ref{fig:experiment}A shows the isometric view of the cells leading-edge. The protrusions are highly dynamic for the round fiber of $D=200~nm$, while less for the flat-ribbon fiber (Fig. \ref{fig:experiment}B). The kymographs in Fig. \ref{fig:experiment}C clearly demonstrate that the ruffles on the round fibers coil around the fiber and move from side to side of the fiber axis, while on the flat-ribbon fibers they stay only on one side and do not wrap around the axis. 

\section{Discussion}

Cells often encounter extra-cellular fibers on which they adhere, spread and migrate. We show that the tendency of the leading-edge of such cells to coil around adhesive fibers can be understood using a model with a minimal set of components: membrane with curved membrane proteins that induce outwards active forces (representing cytoskeletal activity), and adhesion. Within this model the coiling process emerges spontaneously, and is driven by the physics of minimizing the free energy and the active work. This physics-based model predicts that the coiling will be inhibited when the radius of curvature of the fiber is too small, which prevents coiling around flat ribbons due to high membrane bending energy. This prediction is verified in our experiments. Furthermore, the essential role of the curved membrane proteins in our model can explain the reduced coiling observed in cells with reduced amount of such curved and actin-related proteins \cite{Nain-unpublished}.

There are obvious biological consequences for the coiling process. Cells may form better overall grip when coiling on fibrillar structures (i.e., ECM and cellular processes), aiding cell motility and intercellular interactions. For example, cancer cells exhibit enhanced coiling activity on fibers \cite{mukherjee2019cancer,nain2022}. Furthermore, cells migration on tube-like cellular structures during tissue development and organogenesis, as well as during cancer progression, may utilize membrane coiling. One such example is the process of myelination, during which glial cells of the vertebrate central and peripheral nervous systems produce a multi lamellar substance called myelin around  axons, thereby allowing fast nerve conduction \cite{Nave2014,Wilson2021}.  In the peripheral nervous system,  Schwann cells myelination is a process that requires these cells to efficiently coil around fiber-like axons \cite{Gatto2003,Pedraza2009}. In order to closely visualize coiling of Schwann cells on axons, we used a transgenic mouse which expressed a GFP-tagged myelin-specific membrane protein \cite{ERB2006} to generate Schwann cell-neuron myelinating cultures with neurons that expressed TdTomato (red fluorescence) in their cytoplasm. Time lapse imaging of these cultures clearly showed that already at the initial Schwann cell-axon interaction (Fig.6A, Movie-S20) the Schwann cell sent thin processes that coiled around the axons that were to be myelinated. During the myelination process itself (Fig.6B, Movie-S21), we visualized slower and more prominent spiraling of the Schwann cell membrane around the axon, as demonstrated by a kymograph (Fig. 6C), which probably represents wrapping of the inner myelin layer, as known to occur during myelin formation \cite{Snaidero2014}.   It is thus clear that coiling of Schwann cell membranes around axons is a main feature of the unique inter-cellular interactions between Schwann cells and the cylindrical axonal processes.

Our theoretical work indicates that there is a fundamental physical mechanism that leads to membrane coiling dynamics, which can then be further tuned and regulated by biological molecular signaling. Furthermore, myelination is known to occur on axons for which diameter exceeds a critical threshold \cite{Nave2014}. Our results reveal a physical mechanism that inhibits coiling, and may correspondingly prevent myelination around very thin fibers in the vertebrate nervous system.

In conclusion, we propose a physics-based mechanism for the tendency of adhering cellular membranes to coil around fibers. The mechanism is driven by the out-of-equilibrium (active) nature of the forces exerted by the cytoskeleton on the membrane, and the feedback between curved membrane proteins, adhesion and membrane bending. This mechanism may underlie crucial processes in biology, and remains to be tested by more detailed molecular studies.
\begin{figure*}[h!]
\centering
\includegraphics[scale=0.42]{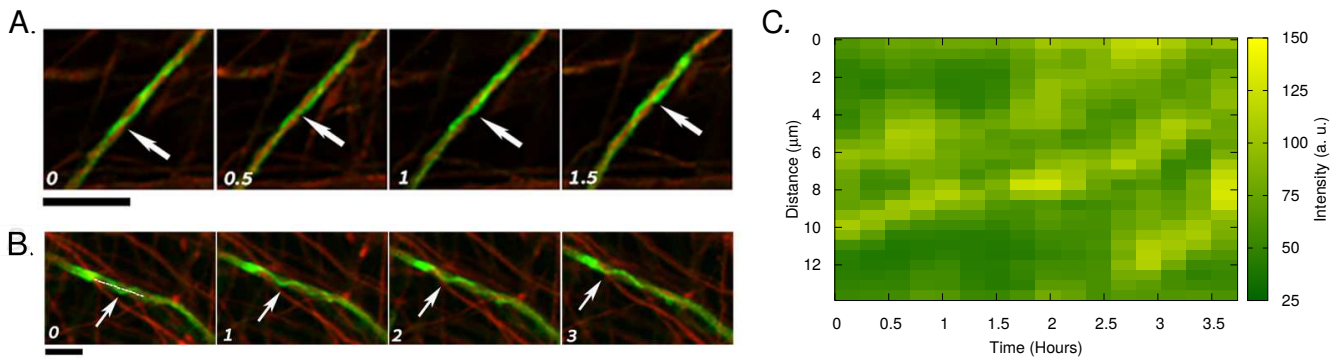}
\caption{Coiling events involved in inter-cellular interactions during vertebrate myelin development. Time lapse imaging of mouse DRG myelinating culture showing (A) initial contact between a thin process of pre-myelinating Schwann cell (green) spiraling around an axon (red). Time intervals between images are $30$ minutes. (B) As myelination proceeds, a thicker Schwann cell process wraps around an axon at a slower speed. Time intervals between images: $1$ hour. Arrows point at clear spiral turns, as they rotate. (C) Kymograph showing fluorescence intensity along a fixed line  across the spiral shown in (B) (depicted as a dashed white line) over the duration of $4$ hours in 15 minute intervals.  Scale bars: $10~\mu m$. The kymograph clearly shows the apparent upwards motion of the fluorescent Schwann cell process, as the spiral shape rotates around the axon.}
\label{myalination}
\end{figure*}

\section{Materials and methods}
\label{sec:model}
\subsection{Theoretical model}
\begin{figure}[h!]
\centering
\includegraphics[scale=0.8]{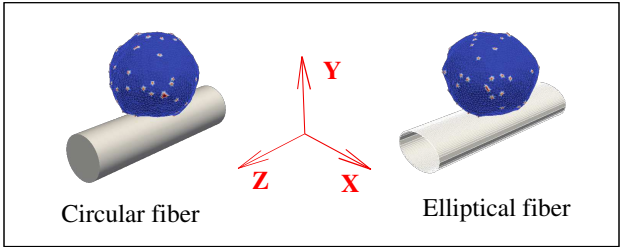}
\caption{Schematic representation of our theoretical model. We consider a three dimensional vesicle, and an adhesive fiber of circular as well as elliptical cross-section. The red dots on the vesicle represent the curved proteins, while the blue regions are the bare membrane.}
\label{fig:model}
\end{figure}

We consider a three-dimensional vesicle, which is described by a closed surface having $N$ vertices, each of them is connected to its neighbors with bonds, and forms a dynamically triangulated, self-avoiding network, with the topology of a sphere \cite{miha2019,Sadhu2021,Sadhu_phagocytosis}. The vesicle is placed on a fiber, with which the vesicle has a uniform attractive contact-interaction (adhesion), as shown in Fig. \ref{fig:model}. The vesicle energy has the four following contributions: The bending energy is given by,

\begin{equation}
    W_b=\frac{\kappa}{2} \int_A (C_1 + C_2 - C_0)^2 dA,
\end{equation}

where $\kappa$ is the bending rigidity, $C_1$, $C_2$ are the principal curvatures and $C_0$ is the spontaneous curvature. We consider the spontaneous curvature $C_0$ to be non-zero at the location of the curved proteins, while in the absence of any curved proteins, the spontaneous curvature is considered to be zero. We consider only convex curved proteins in our simulation, such that their spontaneous curvature $C_0= 1.0 l_{min}^{-1}$. The protein-protein interaction energy is given by, 

\begin{equation}
    W_d = -w\sum_{i<j} {\cal H} (r_0 - r_{ij}),
\end{equation}

where, ${\cal H}$ is the Heaviside step function, $r_{ij}=|\overrightarrow{r}_j - \overrightarrow{r}_i|$ is the distance between proteins and $r_0$ is the range of attraction, $w$ is the strength of attraction. The range of attraction is chosen in a way so that only the proteins that are in the neighboring vertices can interact with each other. The active energy is given by, 

\begin{equation}
    W_F = -F \sum_i \hat{n_i}.\overrightarrow{r}_i,
\end{equation}

where, $F$ is the magnitude of the force, $\hat {n_i}$ is the outward normal vector of the vertex that contains a protein and $\overrightarrow{r}_i$ is the position vector of the protein. These forces are "active" since they give an effective energy (work) term that is unbounded from below and thereby drive the system out-of-equilibrium. If the proteins are distributed inhomogeneously, there will be a net force in a particular direction. However, since we only simulate vesicles that are adhered to the rigid fibers, and the fiber is fixed in its location, this links our vesicle to the lab frame thereby restoring momentum conservation (fixing the fiber acts as an infinite momentum reservoir).

Finally, the adhesion energy, which is due to the attractive interaction between the fiber and the vesicle, is given by, 

\begin{equation}
    W_A = -\sum_{i'} E_{ad},
\end{equation}

where $E_{ad}$ is the adhesion strength, and the sum runs over all the vertices that are adhered to the fiber \cite{samo2015,miha2019,Sadhu2021}. By `adhered vertices', we mean all such vertices, whose perpendicular distance from the surface of the fiber are less than $l_{min}$. Thus, the total energy of the system is given by,

\begin{equation}
    W = W_b + W_d + W_F + W_A
\end{equation}

We update the vesicle with mainly two moves, (1) Vertex movement and (2) Bond flip. In a vertex movement, a vertex is randomly chosen and attempt to move by a random length and direction, with the maximum possible distance of $0.15 l_{min}$. In a bond flip move, a single bond is chosen,  which is a common side of two neighboring triangles, and this bond is cut and reestablished between the other two unconnected vertices \cite{samo2015,miha2019,Sadhu2021}. The maximum bond length is chosen to be $l_{max}=1.7 l_{min}$. The fiber is assumed to be infinite in length in the axial direction ($Z$-axis), while having a finite cross-section (either circular or elliptical) on the $X-Y$ plane (Fig. \ref{fig:model}).

We update the system using Metropolis algorithm, where any movement that increases the energy of the system by an amount $\Delta W$ occurs with rate $exp(- \Delta W/k_BT)$, otherwise it occurs with rate unity. 

We used in this study a vesicle of total number of vertices, $N=3127$ (radius $\sim 20~ l_{min}$), where $l_{min}$ is the unit of length in our model, and defines the minimum bond length. The bending rigidity $\kappa =20 ~k_B T$, the protein-protein attraction strength $w=1 ~k_BT$, and $\rho=N_c/N$ is the protein density, where $N_c$ are the number of vertices occupied by curved membrane proteins with spontaneous curvature: $C_0=1.0~ l_{min}^{-1}$. We do not conserve the vesicle volume here, while the membrane area is well conserved ($\Delta A/A < 1\%$) \cite{miha2019}.

Note that our simulation length scale $l_{min}$ does not have any correspondence with a real length. By assigning a different real length to the basic length-scale of the simulation ($l_{min}$) we do not in fact change the dynamics. The effect of such a change in length scale will be just to rescale the parameters ($E_{ad}$, $F$, $C_0$ etc.) according to the choice of $l_{min}$. The absolute scaling of $l_{min}$ does not therefore make any qualitative change in our simulations, since all the energies are scale-invariant with our definition of $E_{ad}$ and $F$, while the bending energy is always scale-invariant (intensive energy).

\begin{table}[ht]
\begin{center}
\begin{tabular}{|c|c|c|}
\hline
\multicolumn{3}{|c|}{Parameters used in simulation}\\
\hline
Symbol &Definition &Unit\\
\hline
$N$ &Number of vertices &NA\\
\hline
$R$ &Radius of the round fiber &$l_{min}$\\
\hline
$R_x~(R_y)$ &Semi-major (minor) axis of the flat fiber &$l_{min}$\\
\hline
$E_{ad}$ &Adhesion strength &$k_BT$ (per node)\\
\hline
$\rho$ &Density of curved proteins &NA\\
\hline
$w$ &Strength of protein-protein binding&$k_BT$\\
\hline
$F$ &Active protrusive force&$k_BT/l_{min}$\\
\hline
$C_0$ &Spontaneous curvature&$l_{min}^{-1}$\\
\hline
\end{tabular}
\end{center}
\caption{List of parameters used in our simulation}
\label{tab:MC}
\end{table}

%https://www.overleaf.com/project/60e9aa150f8db124653185fe

\subsection{Imaging of the coiling of leading-edge protrusions on suspended fibers}
%------------------------------------------------
\subsubsection{Scaffold Preparation}
Using the previously reported non-electrospinning STEP technique \cite{Nain2009,Nain2013,nain2014}, suspended fiber nanonets composed of 200 or 500 nm diameter fibers spaced 20 μm apart were deposited orthogonally on 2 μm diameter fibers spaced 300 μm apart. Nanofibers were manufactured from solutions of polystyrene (MW: 2,500,000 g/mol; Category No. 1025; Scientific Polymer Products, Ontario, NY, USA) dissolved in xylene (X5-500; Thermo Fisher Scientific, Waltham, MA, USA) in 6 and 9 wt\% solutions. 1 and 3.5 μm diameter fibers were manufactured from 2 and 5 wt\% of high molecular weight polystyrene (MW: 15,000,000 g/mol, Agilent Technologies, Santa Clara, CA, USA) and equally spaced at approximately 60 µm. The polymeric solutions were extruded through micropipettes with an inside diameter of 100μm (Jensen Global, Santa Barbara, CA, USA) for deposition of fibers on a hollow substrate. All fiber networks were crosslinked at intersection points using a custom fusing chamber, to create fixed-fixed boundary conditions. 200nm diameter fibers were made into flat ribbons (of width $~\pi D$, where $D$ is the diameter of the original fiber of circular cross-section) as described in \cite{koons2017}.

\subsubsection{Cell Culturing}
Mouse Muscle Myoblasts (C2C12s) expressing GFP actin were a gift from the Konstantopoulos Lab at Johns Hopkins University. Cells were cultured on Petri dishes using Dulbecco’s Modified Eagle Medium (DMEM, Gibco, Thermo Fisher Scientific) with 10\% Fetal Bovine Serum (FBS, Invitrogen, Carlsbad, CA, USA) and 1\% Penicillin-Streptomycin. STEP-spun scaffolds were placed on 100x20mm Petri dishes and disinfected with 70\% ethanol, then coated with 6µg/ml of Rhodamine Fibronectin (Cat. \# FNR01, CYTOSKELETON, Denver, CO, USA) by incubating at $37^oC$ for 1hr. The trypsinized and resuspended cells were seeded onto STEP-spun scaffolds, allowed to attach for at least two hours and the wells were flooded with 2mL of DMEM +10\% FBS.
\subsubsection{Cell Imaging with Dual-View Light Sheet Microscopy}  
The seeded samples with DMEM were PBS washed, then flooded with Live-Cell Imaging Solution (Invitrogen, Carlsbad, CA, USA) for imaging using Dual-View Plane Illumination Microscopy (diSPIM) \cite{Kumar2014,Wu2013}. Samples were mounted in the diSPIM and kept at $37^oC$ after calibrating the light-sheet movement to piezo step factor using a micromanager plugin \cite{Ardiel2017}. Cells were volumetrically imaged (two views), up to 120 time points at 1.5, 2.5, and 3.5-second intervals constrained by the cell size, the number of slices, and location in the suspended nanonets.
\subsubsection{Cell Imaging with Dual-View Light Sheet Microscopy}
Z stacks acquired from the diSPIM were background subtracted, bleach corrected, cropped, and sorted in Fiji \cite{Schindelin2012}. The orthogonal views (SPIM A and SPIM B) were fused using a GPU optimized pipeline as described in Guo 2020 \cite{Guo2020}. Fused stacks with isotropic resolution were rotated with the TransformJ plugin for visualization and analysis. Maximum intensity projections derived from the volumetric time series were used to manually track the protrusive activity at the coiling fronts of cells. RStudio was used to generate all plots and statistical comparisons of means using Kruskal-Wallis test.  

\subsection{Imaging of early stage myelination process}
For the generation of Schwann cell-dorsal root ganglia (DRG) neuron myelinating cultures we used mice expressing S-MAG-GFP  (a transgene expressed specifically in myelinating cells membranes \cite{ERB2006}). DRG cultures were prepared from mouse embryos at day $13.5$ of gestation. DRGs were dissociated and plated at a density of $4 \times 10^4$ per chamber (Lab-Tek coverglass system Nunc \#155411), coated with Matrigel (Becton Dickinson) and Poly-L-lysine. Cultures were grown for $2$ days in Neurobasal medium supplemented with B-27, glutamax, penicillin /streptomycin and $50~ng/ml$ NGF. At day $2$ post seeding cultures were infected with a lentivirus carrying a cytoplasmic Tdtomato reporter, resulting in neuronal expression of TdTomato. Cultures were then grown for 4 additional days in BN medium containing Basal medium-Eagle, ITS supplement, glutamax, $0.2\%$ BSA, $4~mg/ml$ D-glucose, $50~ng/ml$ NGF and antibiotics. To induce myelination, cultures were grown in BNC, namely a BN medium supplemented with $15\%$ heat inactivated fetal calf serum (replacing the BSA) and $50~\mu g/ml$ L-ascorbic acid.

Cultures were imaged at $7$ days with myelinating medium (for fluorescence imaging, at a frequency of $4$ frames per hour for $65$ or $19$ hours). Fluorescence images were obtained using a confocal microscope (LSM700 confocal microscope Carl Zeiss) $488$-nm and $555$-nm laser lines. Confocal time-lapse images were captured using a Plan-Apochromat $20 \times /0.8$ M27 objective (Carl Zeiss), at temperature , CO2 and humidity controlled conditions. Image capture was performed using acquisition Blocks  (Carl Zeiss Zen 2012). Images and movies were generated and analyzed using Zeiss Zen 2012 and Adobe Photoshop CC 2019.

%\section{Author Contributions}
%Write author contributions here.

\section{Acknowledgment}
ASN acknowledges partial funding support from National Science Foundation (NSF, Grant No. 1762468). ASN and BB acknowledge the Institute of Critical Technologies and Science (ICTAS) and Macromolecules Innovative Institute (MII) at Virginia Tech for their support in conducting this study. HS  acknowledges the support from the intramural research program of the National Institute of Biomedical Imaging and Bioengineering within the National Institutes of Health. The basic version of the code (for theoretical modelling) is provided by Samo Peni\v{c}, which is further edited by RKS for the implementation of the present model. AI acknowledges the  support from the  Slovenian Research Agency (ARRS) through  Programme No. P2-0232 and project No.  No. J3-3066.

\section{Supplementary movies}
All the supplementary movies are available online in the link: https://app.box.com/s/slkec3eu8h34fo6p9s414s0km4a8jgw5

\pagebreak

\setcounter{equation}{0}
\renewcommand{\thefigure}{S-\arabic{figure}}
\renewcommand{\thesection}{S-\arabic{figure}}
\renewcommand{\theequation}{S-\arabic{figure}}
\setcounter{section}{0}
\setcounter{figure}{0}

\section{Supplementary informations}

\subsection{Results for passive vesicle}
\begin{figure}[h!]
\centering
\includegraphics[scale=1]{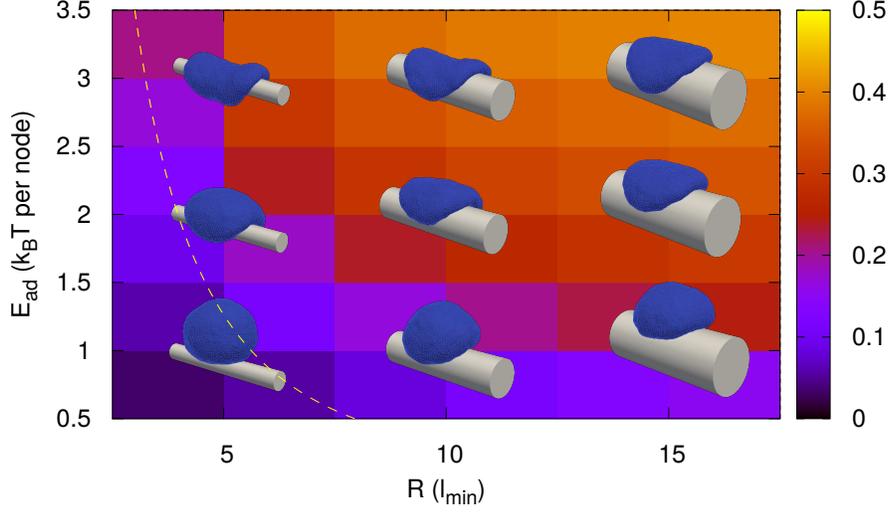}
\caption{Phase diagram $E_{ad}-R$ plane for protein-free case ($\rho=0$). Background color is showing the adhered fraction of vesicle. Adhered fraction increases either with $R$ or $E_{ad}$.}
\label{fig:protein-free}
\end{figure}

We present here the results for the adhesion and spreading of protein-free vesicles on cylindrical substrates, for different fiber radius and strength of adhesion (Fig. \ref{fig:protein-free}). As expected, the adhered area fraction increases strongly with both fiber radius and adhesion energy, with the transition to complete adhesion determined by the balance between the bending and adhesion energies. At large adhered area fractions, the bending energy per vesicle-substrate area balances the adhesion area, with the complete adhesion regime occurring when: $\kappa/R^2\leq E_{ad}$, shown by the yellow dashed line in Fig. \ref{fig:protein-free}. 
%%%%%%%%%%%%%%%%%%%%%%%%%%%%%%%%%%%%%%%%%%%%%%%%%%%%%%%%%%%%%%%%%%%%%%%%%%%%%%%%%%

\begin{figure}[h!]
\centering
\includegraphics[scale=1.5]{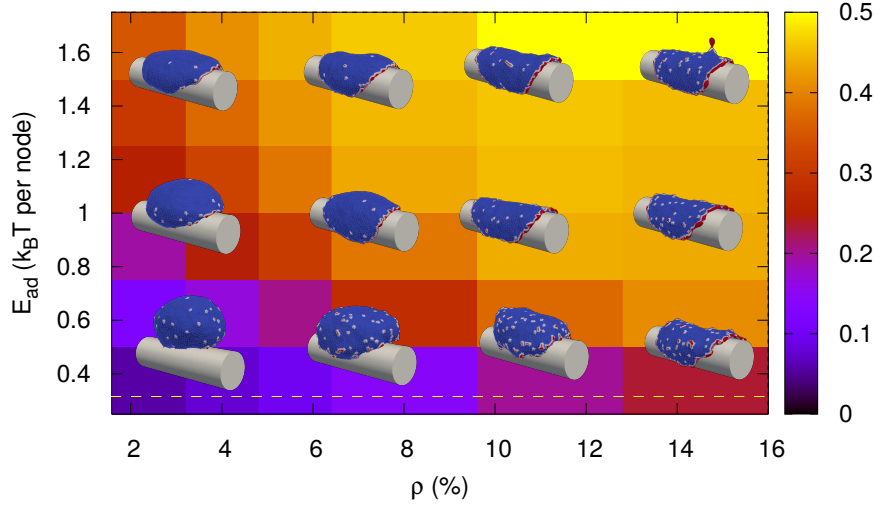}
\caption{Phase diagram $E_{ad}-\rho$ plane for vesicle with passive proteins ($F=0$). Background color is showing the adhered area fraction. The adhered fraction increases with $E_{ad}$ and $\rho$. }
\label{fig:passive-proteins}
\end{figure}

We next show our results with passive curved proteins ($F=0$) and a given radius of fiber $R=10~l_{min}$ in Fig. \ref{fig:passive-proteins}. With the introduction of curved passive proteins, the complete adhesion regime is extended to lower adhesion energies, as was observed for flat substrates \cite{Sadhu2021}. For a given small $E_{ad}$, as we increase protein density $\rho$, the vesicle gets adhered more and more. For larger $E_{ad}$, however, the strong adhesion is achieved even without (or small) proteins. The yellow horizontal dashed line is showing the transition given by $\kappa/R^2\leq E_{ad}$ (for a given $R$) above which the adhesion energy dominates.

%%%%%%%%%%%%%%%%%%%%%%%%%%%%%%%%%%%%%%%%%%%%%%%%%%%%%%%%%%%%%%%%%%%%%%%%%%%%%%%%%%%
\subsection{Analytical estimation of transition lines}
Here, we present our simple analytical estimation of all the transition lines separating different phases (shown in Fig. 2d, main text). 
%%%%%%%%%%%%%%%%%%%%%%%%%%%%%%%%%%%%%%%%%%%%%%
\subsubsection{Transition to a pancake shape (phase-I to II or  phase-IV to III)}
Let us consider the surface area of the vesicle is $A$. If we make this a flat pancake-like, the radius of this circular face will be given by the relation,

\begin{equation}
    \pi R_p^2 \simeq A/2
\end{equation}

or, 

\begin{equation}
    R_p \simeq \sqrt{A/2\pi}
\end{equation}

The perimeter of the circular face of the pancake-like shape is,

\begin{equation}
    2 \pi R \simeq \sqrt{2 \pi A}
\end{equation}

If the average size of a protein is, $l_{av}=(l_{min} + l_{max})/2$, then, the number of proteins that will fit in that perimeter is given by,

\begin{equation}
    N_p \simeq \frac{\sqrt{2 \pi A}}{l_{av}}
\end{equation}

Now, the average surface area of the vesicle is,

\begin{equation}
    A=2 N \frac{\sqrt{3}}{4}l_{av}^2
\end{equation}

where, $N$ is the number of nodes, and $2N$ is the approximate number of triangles. Since, $N=3127$, and $l_{av}=1.35~l_{min}$, the total number of proteins that will fit into the rim of circular pancake is,
 
\begin{equation}
    N_p=\frac{\sqrt{2 \pi A}}{l_{av}} \simeq 130
\end{equation}
 
 The density of proteins is thus,
 
 \begin{equation}
    \rho_p (\%)=(100 \times 130)/N \simeq 4 \%
\end{equation}
 
%%%%%%%%%%%%%%%%%%%%%%%%%%%%%%%%%%%%%%%%%%%
\subsubsection{Transition from ``hanging'' pancake (phase-II) to ``wrapped'' pancake (phase-III)}
Let us assume the vesicle forms a circular pancake-like shape with the radius of $R_p$, such that the total area of the vesicle, $A=2 \pi R_p^2$. Now, consider the case where the pancake just touching the cylinder and remains unadhered. The adhesion energy in this case is,

\begin{equation}
    W_A = -E_{ad} \frac{2 d R_p}{l^2_{av}}
\end{equation}

Here, $l_{av}$ is the average length of the bond, $l_{av}=(l_{min} + l_{max})/2$. Here, we assume that the width that is in contact with the cylinder is of the length $d=l_{min}$, such that the number of adhered nodes are $2R_p/l^2_{av}$. After the pancake is fully adhered, the new adhesion energy and the bending energy are,

\begin{equation}
    W_A = - E_{ad} \frac{\pi R_p^2}{l^2_{av}}
\end{equation}

and the extra bending energy cost for this transition,

\begin{equation}
    W_b = \kappa \pi R_p^2 \frac{1}{R^2}
\end{equation}

The condition for the transition is,

\begin{equation}
    \frac{E_{ad}}{l^2_{av}} (\pi R_p - 2) = \kappa \frac{\pi R_p}{R^2}
\end{equation}

which gives,

\begin{equation}
    R^2 = \frac{\kappa}{E_{ad}} \frac{l^2_{av}}{\left( 1 - \frac{2}{\pi R_p}  \right)}
\end{equation}

Assuming $R_p \gg 1$, we have,

\begin{equation}
    R \sim \sqrt{\kappa / E_{ad}} l_{av} \sim 6 l_{min}
\end{equation}

%\begin{figure}[H]
%\centering
%\includegraphics[scale=0.7]{analytical_transiton_compare.eps}
%\caption{Comparison of analytical prediction of transition lines with the simulation. The dashed lines with points are the simulation results and the solid lines are the analytical predictions. }
%\end{figure}
%%%%%%%%%%%%%%%%%%%%%%%%%%%%%%%%%%%%%%%%%%%%
\subsubsection{Coiling transition (phase-I to phase-IV)}
\begin{figure}[h!]
\centering
\includegraphics[scale=0.8]{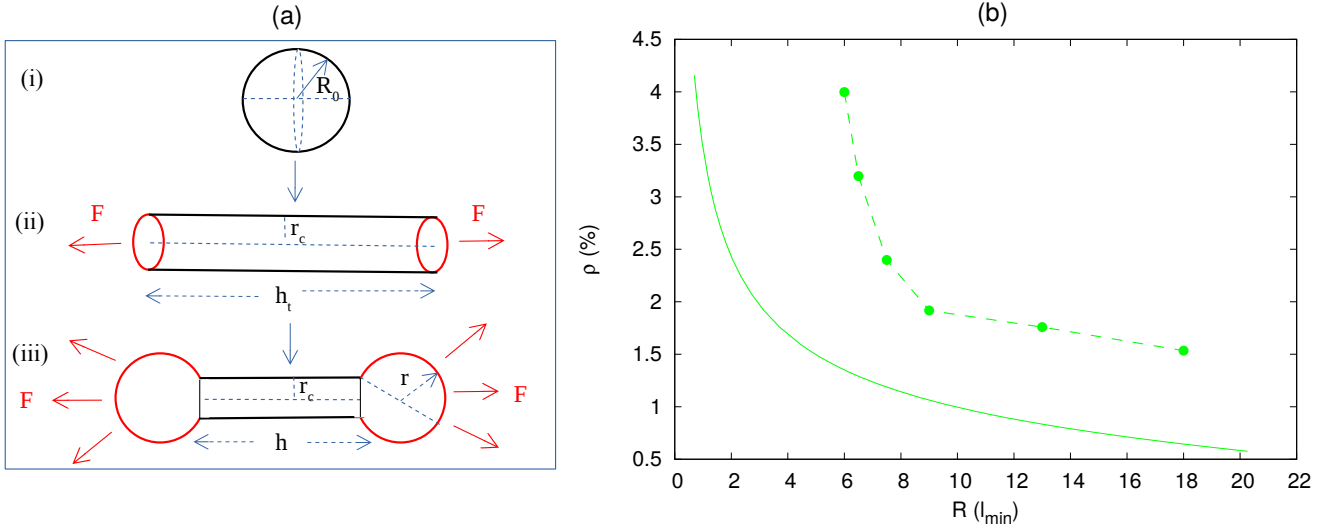}
\caption{Analytical estimation of transition line for coiling transition (phase-I to phase-IV). (a) Schematic representation of steps of analytical calculations. (i) For the vesicle without any force, the shape is spherical. (ii) For small $R$, when force is applied, the vesicle shape changes to cylindrical one. (iii) For larger $R$, the vesicle shape changes to coiling shape, having a combination of cylindrical and circular shape. The red lines are representing the curved proteins. (b) Comparison of analytical estimation with the simulation. Other simulation parameters are same as in Fig. 2, main text.}
\label{fig:analytical-coiling}
\end{figure}

For small $\rho$, the vesicle either forms elongated shapes axially aligned, or coiled shape. For small $R$, the vesicle aligned along the axial direction and for large $R$, it forms a circumferential (coiled) shape. Let us assume a spherical vesicle having zero force (Fig. \ref{fig:analytical-coiling}a(i)). As the force increases, the vesicle will get stretched (Fig. \ref{fig:analytical-coiling}a(ii)). Using the energy minimization, and area conservation, we can get the elongated shape, assuming it to be cylindrical.
The area conservation states,

\begin{equation}
    4 \pi R_0^2 \simeq 2 \pi r_c h_t
\end{equation}

where, we neglect the area for the two circular edges. Now, assume that for spherical vesicle, only a single vertex of adhered, while for cylindrical shape, a linear chain of vertices are adhered along the axial direction. Thus, from energy minimization,

\begin{equation}
    (N_c R_0 F + E_{ad} - 8 \pi \kappa) r_C^2 - 2 R_0^2 (E_{ad}/l_{av} + N_c F/2) - 8 \pi \kappa R_0^2 (N_c R_0 F + E_{ad} - 8 \pi \kappa) = 0
\end{equation}

The solution of this equation will give two values of $r_c$ among which, one is greater than $R_0$ and hence is unphysical. We consider the other solution We can also calculate $h_t$ from area conservation.

Now, let us assume that for large $R$, when the vesicle goes to a coiled shape, the value of $r_c$  remains same, and the shape becomes a combination of cylinder and two flat circular shapes (Fig. \ref{fig:analytical-coiling}a(iii)). The radius of the flat circular shapes will be determined by the protein density,

\begin{equation}
    2 \pi r - 2 r_c = N_c l_{av}/2
\end{equation}

or,

\begin{equation}
    r = (N_c l_{av}/2 + 2 r_c)/2 \pi
\end{equation}

From area conservation,

\begin{equation}
    4 \pi R_0^2 = 2 \pi h r_c + 4 \pi r^2
\end{equation}

and from energy minimization,

\begin{equation}
    -E_{ad} h_t/l_{av} + \frac{\kappa}{2} 2 \pi r_c h_t \frac{1}{r_c^2} = -E_{ad} (h/l_{av} + 2 \pi r^2/l_{av}^2) + \frac{\kappa}{2} 2 \pi r_c h \frac{1}{r_c^2} + \frac{\kappa}{2} 4 \pi R_0^2 \frac{1}{R^2}
\end{equation}

\begin{equation}
    R^2 = \frac{\kappa A}{2} [ E_{ad} (h/l_{av} + 2 \pi r^2/l_{av}^2 - h_t/l_{av}) + \pi \kappa/r_c (h_t-h) ]^{-1}
\end{equation}

where, $R$ is the radius of the cylinder. We solve the above equation numerically in the $R-\rho$ plane, and show in Fig. \ref{fig:analytical-coiling}(b) with green solid line, and compare with simulation results with dashed line-points.

%%%%%%%%%%%%%%%%%%%%%%%%%%%%%%%%%%%%%%%%%%%%%%%%%%%%%%%%%%%%%%%%%%%%%%%%%%%%%%%%%%%
\subsection{Quantification of axial and circumferential alignment}
\begin{figure}[h!]
\centering
\includegraphics[scale=1]{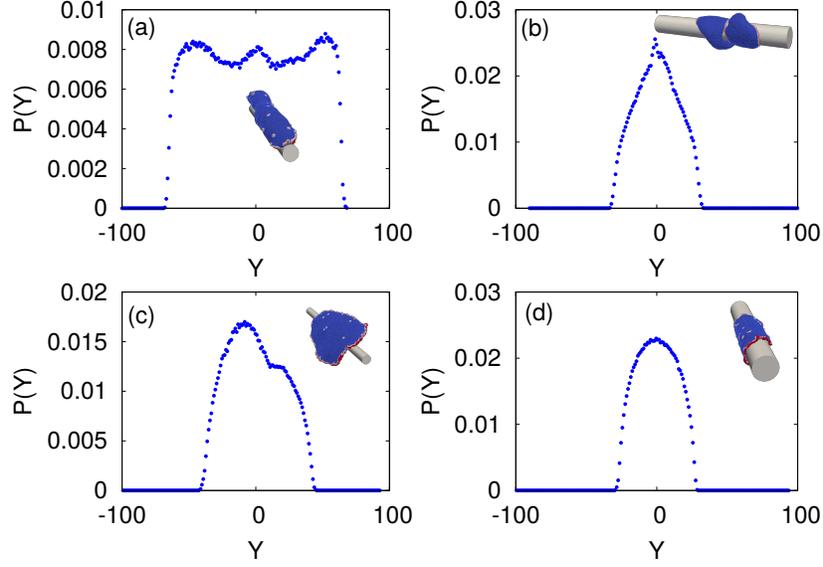}
\caption{Distribution of vertices along the axial direction. (a) $\rho=3.2 \%$ and $R=3.0 l_{min}$. (b) $\rho=3.2 \%$ and $R=8.0 l_{min}$. (a) $\rho=9.6 \%$ and $R=3.0 l_{min}$. (a) $\rho=9.6 \%$ and $R=8.0 l_{min}$.  }
\label{fig:axial}
\end{figure}

\begin{figure}[h!]
\centering
\includegraphics[scale=1]{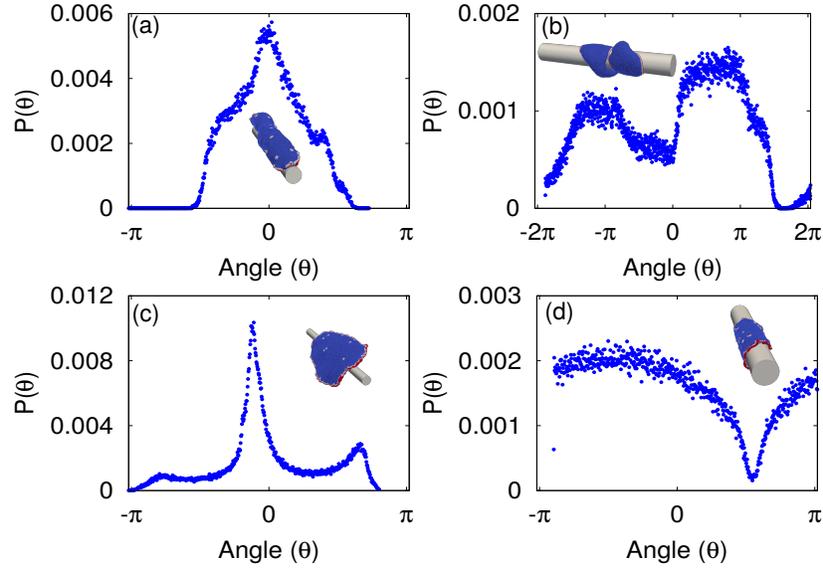}
\caption{Angular distribution of the vertices along the circumferential direction. (a) $\rho=3.2 \%$ and $R=3.0 l_{min}$. (b) $\rho=3.2 \%$ and $R=8.0 l_{min}$. (a) $\rho=9.6 \%$ and $R=3.0 l_{min}$. (a) $\rho=9.6 \%$ and $R=8.0 l_{min}$.  }
\label{fig:circumferential}
\end{figure}

In our main text, we quantify the axial and circumferential alignment of the vesicle by measuring the variance in the distribution of vertices along the axial direction and the distribution of angle along the circumferential direction respectively. Here, we show the full distribution of the vertices along the axial and circumferential direction in Fig. \ref{fig:axial} and Fig. \ref{fig:circumferential} respectively.

%%%%%%%%%%%%%%%%%%%%%%%%%%%%%%%%%%%%%%%%%%%%%%%%%%%%%%%%%%%%%%%%%%%%%%%%%%%%%%%%%%%
\subsection{Reorientation process: Energy with time without scaling per vertex}
\begin{figure}[h!]
\centering
\includegraphics[scale=1.4]{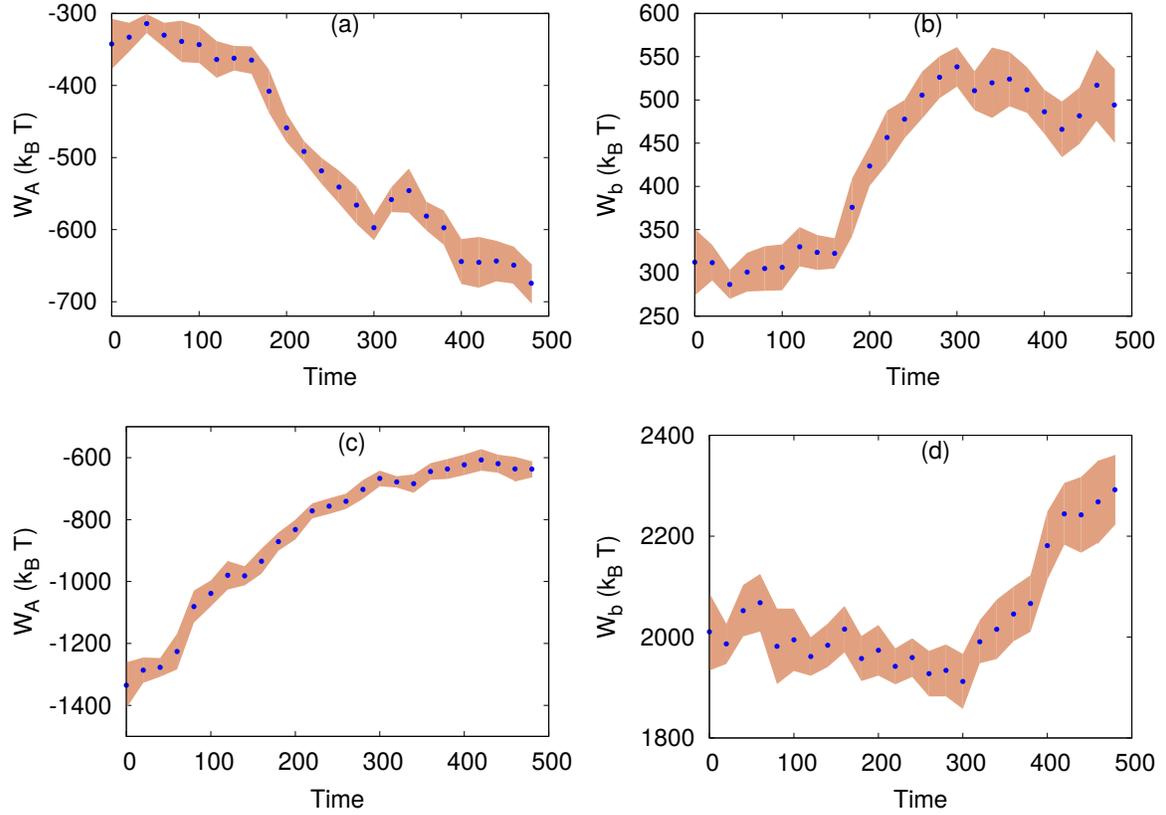}
\caption{(a) Adhesion energy with time for an arc. (b) Bending energy with time for an arc. (c) Adhesion energy with time for the rest cylindrical part. (d) Bending energy with time for the rest cylindrical part. Other simulation parameters are same as in Fig. 3, main text.}
\label{Fig:energy-no-scaling}
\end{figure}

Here, we show the adhesion and bending energies for an arc without scaling by the total number of vertices that are forming the arc, in Fig. \ref{Fig:energy-no-scaling}(a-b). We also show the energy of the rest of the cylindrical part of the vesicle (after subtracting the contribution due to the other two arcs) in Fig. \ref{Fig:energy-no-scaling}(c-d). We note that the adhesion energy is increasing throughout and bending energy is also larger in the final configuration.

%%%%%%%%%%%%%%%%%%%%%%%%%%%%%%%%%%%%%%%%%%%%%%%%%%%%%%%%%%%%%%%%%%%%%%%%%%%%%%%%%%%
\subsection{Coiling speed for different fiber radius}
\begin{figure}[h!]
\centering
\includegraphics[scale=0.7]{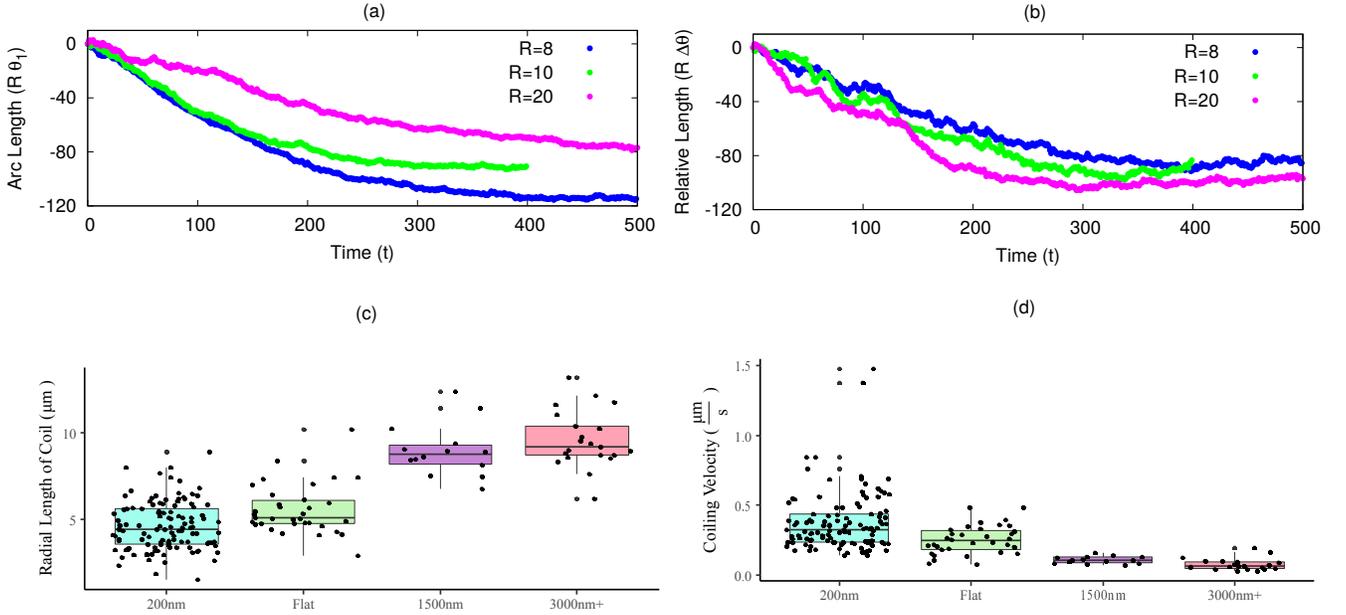}
\caption{Coiling speed for different fiber radius in simulations, and compared with experiments. (a) The overall angular-length-span of the vesicle coil as function of time. (b) The displacements of the leading-edge arcs along the angular direction, as function of time. (c) Radial length of coil trajectory measured from experiments. (d) Coiling velocity from the experimental data, for each of the four cases. }
\label{fig:coiling-velocity}
\end{figure}

In Fig. \ref{fig:coiling-velocity} we compare the coiling velocity of the leading edge protrusion measured from experiments with the simulation data. The simulation data indicates that over the fiber radii that we can test, the coiling speed does not change significantly (slopes of the graphs in Fig. \ref{fig:coiling-velocity}b). However, in the experiments the coiling speed was found to decrease for increasing radii, most significantly for small radius (Fig. \ref{fig:coiling-velocity}d).

%%%%%%%%%%%%%%%%%%%%%%%%%%%%%%%%%%%%%%%%%%%%%%%%%%%%%%%%%%%%%%%%%%%%%%%%%%%%%%%%%%%%%%%

\section{Supplementary movies}
All the supplementary movies are available online in the link: https://app.box.com/s/slkec3eu8h34fo6p9s414s0km4a8jgw5 \\

\begin{itemize}
\item \textbf{Movie-S1}{Experimental movie corresponding to Fig. 1A. Deconvolved isometric render in ImageJ 3D Viewer of a C2C12 expressing GFP actin spread on a rhodamine fibronectin coated $200~nm$ fiber. Scale bar shown is for $10 \mu m$.}
\label{movie1}

\item \textbf{Movie-S2}{Experimental movie corresponding to Fig. 1B. Zoomed-in perspective of leading/trailing edge of the cell as it migrates and coils around the 200nm fiber. Scale bar shown is for $5 \mu m$}
\label{movie2}

\item \textbf{Movie-S3}{Experimental movie corresponding to Fig. 1C. Zoomed-in perspective of leading/trailing edge of the cell as it migrates and coils around the 200nm fiber. Scale bar shown is for $5 \mu m$.}
\label{movie3}

\item \textbf{Movie-S4}{Spreading of protein-free vesicle on cylindrical fiber of smaller radius. The parameter values are : $E_{ad}=1.5~k_BT$, $R=5.0~l_{min}$.}
\label{movie4}

\item \textbf{Movie-S5}{Spreading of protein-free vesicle on cylindrical fiber of larger radius. The parameter values are : $E_{ad}=1.5~k_BT$, $R=15.0~l_{min}$.}
\label{movie5}

\item \textbf{Movie-S6}{Spreading of a vesicle with passive proteins (low density) on cylindrical fiber. The parameters are: $E_{ad}=1.0~k_BT$, $\rho=1.6\%$, $R=10.0~l_{min}$.}
\label{movie6}

\item \textbf{Movie-S7}{Spreading of a vesicle with passive proteins (medium density) on cylindrical substrate. The parameters are: $E_{ad}=1.0~k_BT$, $\rho=3.2\%$, $R=10.0~l_{min}$.}
\label{movie7}

\item \textbf{Movie-S8}{Spreading of a vesicle with passive proteins (high density) on cylindrical substrate. The parameters are: $E_{ad}=1.0~k_BT$, $\rho=6.4\%$, $R=10.0~l_{min}$.}
\label{movie8}

\item \textbf{Movie-S9}{Spreading of a vesicle with active proteins with small $R$ and small $\rho$ (Phase-I). The parameters are: $E_{ad}=1.0~k_BT$, $\rho=3.2\%$, $R=4.0~l_{min}$, and $F=2.0~k_BT/l_{min}$.}
\label{movie9}

\item \textbf{Movie-S10}{Spreading of a vesicle with active proteins with small $R$ and large $\rho$ (Phase-II). The parameters are: $E_{ad}=1.0~k_BT$, $\rho=6.4\%$, $R=4.0~l_{min}$, and $F=2.0~k_BT/l_{min}$.}
\label{movie10}

\item \textbf{Movie-S11}{ Spreading of a vesicle with active proteins with large $R$ and large $\rho$ (Phase-III). The parameters are: $E_{ad}=1.0~k_BT$, $\rho=6.4\%$, $R=10.0~l_{min}$, and $F=2.0~k_BT/l_{min}$.}
\label{movie11}

\item \textbf{Movie-S12}{Spreading of a vesicle with active proteins with large $R$ and small $\rho$ (Phase-IV). The parameters are: $E_{ad}=2.0~k_BT$, $\rho=2.4\%$, $R=10.0~l_{min}$, and $F=2.0~k_BT/l_{min}$.}
\label{movie12}

\item \textbf{Movie-S13}{Axial to circumferential (coiling) transition. The parameters are: $E_{ad}=1.0~k_BT$, $\rho=2.4\%$, $R=10.0~l_{min}$, and $F=2.0~k_BT/l_{min}$.}
\label{movie13}

\item \textbf{Movie-S14}{Spreading of vesicle with active proteins on a cylinder with elliptical cross-section. The parameters are: $E_{ad}=1.0~k_BT$, $\rho=3.2\%$, $R_x=12.0~l_{min}$, $R_y=7.77~l_{min}$, and $F=2.0~k_BT/l_{min}$.}
\label{movie14}

\item \textbf{Movie-S15}{Experimental movie corresponding to Fig. 5a, 200nm, leading-edge 1.  Maximum intensity projection of the leading-edge volume used for rotational analysis. Dots and lines are overlayed paths using ImageJ manual tracking plugin.}
\label{movie15}

\item \textbf{Movie-S16}{Experimental movie corresponding to Fig. 5a, 200nm, leading-edge 2. Maximum intensity projection of the leading-edge volume used for rotational analysis. Dots and lines are overlayed paths using ImageJ manual tracking plugin. }
\label{movie16}

\item \textbf{Movie-S17}{Experimental movie corresponding to Fig. 5b, flat ribbon. Maximum intensity projection of the leading-edge volume used for rotational analysis. Dots and lines are overlayed paths using ImageJ manual tracking plugin.}
\label{movie17}

\item \textbf{Movie-S18}{Experimental movie corresponding to Fig. 5c, $1500~nm$. Maximum intensity projection of the leading-edge volume used for rotational analysis. Dots and lines are overlayed paths using ImageJ manual tracking plugin.}
\label{movie18}

\item \textbf{Movie-S19}{Experimental movie corresponding to Fig. 5d, $3000~nm$. Maximum intensity projection of the leading-edge volume used for rotational analysis. Dots and lines are overlayed paths using ImageJ manual tracking plugin. }
\label{movie19}

\item \textbf{Movie-S20}{Time lapse imaging of mouse DRG myelinating culture showing initial contact (arrow) between a pre-myelinating Schwann cell (green) and an axon (red).Images were taken at time intervals of 15min. The Schwann cell process spirals around the axon at a speed of aprx $90$ minutes per round . Figure 6A depicts frames \#68,70,72,74 of this movie. }
\label{movie20}

\item \textbf{Movie-S21}{Time lapse imaging of mouse DRG myelinating culture, taken time intervals of $15$ minutes, showing a myelinating Schwann cell membrane (green, arrow) wrapping around an axon (red) at a speed of aprx $180$ minutes per round, generating a prominent dynamic spiral. Figure 6B depicts frames 52,56,60,64 of this movie.}
\label{movie21}

\end{itemize}

\end{document}